\documentclass[aps,prb,twocolumn,superscriptaddress,floatfix,amsmath,amssymb,showpacs]{revtex4-1}

\usepackage[dvips]{graphicx}
\usepackage[dvipdfm]{hyperref}

\usepackage{epstopdf}
\begin{document}

\title{Magnetic properties of the S=1/2 quasi square lattice antiferromagnet CuF$_2$(H$_2$O)$_2$(pyz) (pyz=pyrazine) investigated by neutron scattering}

\author{C. H. Wang}
\author{M. D. Lumsden}
\author{R. S. Fishman}
\author{G. Ehlers}
\author{T. Hong}
\author{W. Tian}
\author{H. Cao}
\author{A. Podlesnyak}
\affiliation{Oak Ridge National Laboratory, Oak Ridge, Tennessee 37831, USA}
\author{C. Dunmars}
\author{J. A. Schlueter}
\affiliation{Material Science Division, Argonne National Laboratory, Argonne, Illinois, 60439}

\author{J. L. Manson}
\affiliation{Department of Chemistry and Biochemistry, Eastern Washington University, Cheney, Washington 99004}

\author{A. D. Christianson}
\affiliation{Oak Ridge National Laboratory, Oak Ridge, Tennessee 37831, USA}

\date{\today}

\begin{abstract}

We have performed elastic and inelastic neutron scattering experiments on single crystal samples of the coordination polymer compound
CuF$_2$(H$_2$O)$_2$(pyz) (pyz=pyrazine) to study the magnetic structure and excitations. The elastic neutron diffraction measurements indicate
a collinear antiferromagnetic structure with moments oriented along the [0.7 0 1] real-space direction
and an ordered moment of 0.60 $\pm$ 0.03 $\mu_B$/Cu. This value is significantly smaller than the
single ion magnetic moment, reflecting the presence of strong quantum fluctuations.
The spin wave dispersion from magnetic zone center to the zone boundary points (0.5 1.5 0) and (0.5 0 1.5)
can be described by a two dimensional Heisenberg model with a nearest neighbor magnetic exchange constant
$J_{2d}$ = 0.934 $\pm$ 0.0025 meV. The inter-layer interaction
$J_{perp}$ in this compound is less than 1.5\% of $J_{2d}$. The spin excitation energy at the (0.5 0.5 0.5) zone
boundary point is reduced when compared to the (0.5 1 0.5) zone boundary point by $\sim$ 10.3 $\pm$ 1.4 \%. This
zone boundary dispersion is consistent with quantum Monte Carlo and series expansion calculations for the S=1/2 Heisenberg square lattice antiferromagnet which include corrections
for quantum fluctuations to linear spin wave theory.

\end{abstract}


\maketitle

\section{Introduction}

Metal-organic systems with a 3d$^9$ electron configuration, such as in Cu$^{2+}$, are expected to undergo a
Jahn-Teller distortion. In octahedral coordination, this typically elongates one axis of the octahedron  and removes the
degeneracy of the e$_g$ orbitals d$_{x^2-y^2}$ and d$_{z^2}$. This effect widely occurs in molecular systems
as well as in other materials such as the colossal magnetoresistance manganites.\cite{CMR1,CMR2} The
Jahn-Teller distortion is extremely sensitive to bond distances and as such applied pressure can strongly
influence or even induce a Jahn-Teller distortion. Depending on the elongation axis in the crystal
structure, the pattern of orbital overlaps and related exchange interactions can vary significantly in turn leading to disparate
magnetic ground states. Finding materials where a Jahn-Teller distortion can be tuned to act as a
magnetic switch has potential for applications in technological devices and, thus, is of great interest.

The copper-based coordination polymer magnet CuF$_2$(H$_2$O)$_2$(pyz) (pyz=pyrazine) appears to
be a model material for studying the switching of magnetic properties due to changes in the
Jahn-Teller axis.\cite{Schlueterpressure,prescimone2012} CuF$_2$(H$_2$O)$_2$(pyz) crystallizes in a monoclinic
structure (space group P2$_1$/c) with a=7.6926 {\AA} , b=7.5568{\AA} , c=6.897{\AA} and
$\beta$=111.065$^o$ under ambient conditions.\cite{CuF2Manson,CuF2Schlueter} The structure
consists of CuF$_2$O$_2$N$_2$ octahedra which form a 2d network in the bc plane. The
magnetic susceptibility shows a broad peak at 10.5 K consistent with two dimensional (2d)
short range correlations and three dimensional long-range order below T$_N$ = 2.6 K due to weak coupling
along the a-axis. At ambient pressure, the Jahn-Teller axis is
along the N-Cu-N axis (a direction).  Hence the d$_{x^2-y^2}$ magnetic orbital lies in
the bc plane to form a two dimensional (2d) antiferromagnetic (AFM) quasi square lattice where the near neighbor distances are the same, but the interior angles deviate by $\pm$ $\sim$5$^{\circ}$. First principles
electronic structure calculations confirm this configuration.\cite{CuF2Manson} These calculations give an intra-plane AFM
exchange interaction of about 13 to 19 K and an inter-plane AFM exchange interaction of only 1\% of the
intra-plane exchange interaction. An important feature of CuF$_2$(H$_2$O)$_2$(pyz) is that under an
applied pressure of 0.9 GPa, a switch of the Jahn-Teller axis from the N-Cu-N to the O-Cu-O (c direction)
bond occurs.\cite{Schlueterpressure} When pressure increases to 3.1 Gpa, the Jahn-Teller axis switches
again from O-Cu-O bond to F-Cu-F bond (b direction). Correspondingly the magnetic interactions
are expected to vary at different pressures: from the ambient pressure 2d quasi square lattice (exchange path
Cu-F$\cdot \cdot \cdot$H-O-Cu) to one dimensional (1d) chain interactions (exchange path Cu-pyz-Cu) at 0.9 GPa. Indeed, the magnetic
susceptibility at ambient pressure can be well described by the 2 d Heisenberg square lattice \cite{CuF2Manson} while above 0.9 GPa the magnetic susceptibility exhibits a broad peak consistent with 1d magnetic correlations\cite{Schlueterpressure}. Therefore a microscopic understanding of the magnetic
properties at ambient pressure would
be an illuminating step towards a complete understanding of the interesting pressure dependent behavior of CuF$_2$(H$_2$O)$_2$(pyz).

In this paper we present the results of a study of deuterated CuF$_2$(H$_2$O)$_2$(pyz) (CuF$_2$(D$_2$O)$_2$(d$_4$-pyz))
with elastic and inelastic neutron scattering under ambient pressure. The neutron diffraction
results show that the sample has a collinear AFM structure with the moments lying in
ac-plane along the [0.7 0 1] direction. The spin wave dispersion extracted from the inelastic neutron
spectra can be well described by a nearest-neighbor 2d Heisenberg model.
The spin wave dispersion found along the inter-layer direction could not be observed within the
instrumental resolution of FWHM=0.034 meV.  This indicates that the inter-layer exchange interaction
$J_{perp}$ is weak as expected for a 2d system where the spin wave dispersion does not depend on the out of plane direction.  Consequently, CuF$_2$(H$_2$O)$_2$(pyz) is a good example of quasi-2d system where the spin wave dispersion depends exclusively on K and L.
Within the 2d magnetic plane, the dispersion along the zone boundary points has been measured and
is in accord with prediction of the quantum Monte Carlo and series expansion calculations\cite{Singh,expansion2,MontCarlo2} for quantum fluctuation corrections to linear spin wave theory for the S=1/2 square lattice Heisenberg antiferromagnet.

\section{Experimental Details}

In order to minimize attenuation of the beam caused by the large incoherent scattering
cross-section of hydrogen, fully deuterated
CuF$_2$(D$_2$O)$_2$(d$_4$-pyz) crystals were synthesized for this study. Substitution
of deuterium for hydrogen does not alter the crystal symmetry and induces only small
changes in the lattice parameters.\cite{CuF2Schlueter} To ensure that structural parameters of the samples studied here are in accord with previously published values\cite{CuF2Schlueter} neutron diffraction data (not shown) were
collected on the four circle neutron diffractometer HB-3A at the High Flux Isotope Reactor (HFIR), Oak Ridge National Laboratory.  The magnetic structure was studied with the thermal triple-axis
spectrometer HB-1 at the HFIR. A crystal with a mass of 0.01 g was studied using HB-3A while a larger crystal
with mass 0.1 g was studied with HB-1. A silicon monochromator and no analyzer were used for the measurements on HB-3A and unless otherwise noted all triple-axis measurements were performed using pyrolytic graphite (002) monochromator and analyzer crystals.

Inelastic neutron scattering experiments were performed using the cold neutron
triple-axis spectrometer CG-4C (HFIR), the thermal triple-axis HB-1A (HFIR) and the cold neutron
chopper spectrometer CNCS\cite{cncs} at the Spallation Neutron Source (SNS), Oak Ridge National Laboratory. The inelastic neutron
scattering experiments were performed on four co-aligned crystals with total mass of
0.85 g and total mosaic of 0.7$^o$. For the CG-4C experiment, measurements were made with fixed final energy, $E_f$, of 5 meV and 3.7 meV resulting in an elastic energy resolution of about
0.3 meV and 0.17 meV, respectively.
For the HB-1A experiment, the incident energy $E_i$ was fixed at 14.7 meV with a pyrolytic graphite (002) monochromator and a beryllium (002)
analyzer was used to obtain an elastic energy resolution of about 0.58 meV.
For the CNCS measurements, the incident energy $E_i$ was fixed at 3 meV and 1.5 meV to allow examination of the entire spin wave spectrum. The energy resolution (FWHM) was about
0.08 meV and 0.034 meV at the elastic position for $E_i$ = 3 meV and 1.5 meV, respectively.

\section{Results and Discussion}

\subsection{magnetic structure}

\begin{figure}[t]
\centering
\includegraphics[width=0.4\textwidth]{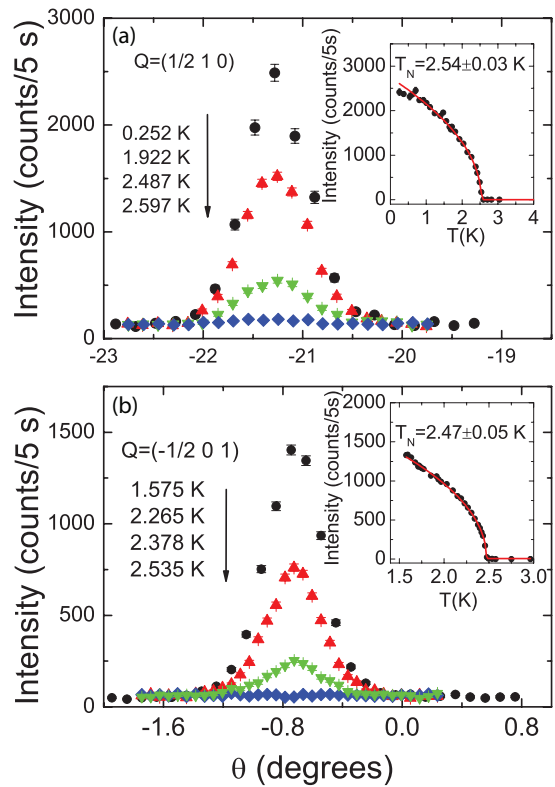}
\caption{\label{fig1} (a) Temperature dependence of magnetic Bragg peaks (a): (0.5 1 0) and
(b): (-0.5 0 1) for CuF$_2$(D$_2$O)$_2$(d$_4$-pyz). Insets display the temperature dependent peak intensity. Solid lines are guides to the eye. The data were collected on the
HB-1 triple-axis instrument in $H$$K$0 and $H$0$L$ scattering planes, respectively.}
\vspace*{-3.5mm}
\end{figure}

Temperature dependent neutron diffraction reveal the presence of additional peaks for temperatures less than 2.5 K  at wave vectors Q=(0.5 $K$ $L$) with integer values of $K$ and $L$ and where $K$+$L$ is
an odd number. This
indicates that in real space the magnetic lattice is doubled along the a direction
compared to nuclear lattice while in the bc-plane the nuclear lattice and magnetic lattice
remain the same size. Fig. 1 shows the temperature dependent magnetic Bragg peaks
at both (0.5 1 0) and (-0.5 0 1) collected on the HB-1 instrument
using the final energy $E_f$=13.5 meV($\lambda$=2.4617 \AA). In the insets the temperature dependent
peak intensity for (0.5 1 0) and (-0.5 0 1) are presented. A power-law fit of the form (1-T/T$_N$)$^{2 \beta}$ gives a
$T_N$ of 2.51 $\pm$ 0.06 K and a value of $\beta$ of 0.24 $\pm$ 0.07.  This low value of $\beta$ is likely consistent with 2d critical behavior, however, care should be used when interpreting this finding as a more rigorous series of measurements is required to definitively establish the critical exponents.  The value of T$_N$ determined from the powerlaw fit is slightly smaller than the value reported by
muon spin relaxation measurements of a non-deuterated sample\cite{CuF2Manson,Goddard} but is in reasonable agreement with the value of $T_N$=
2.53 K  determined from  magnetic susceptibility measurements of fully deuterated
samples.\cite{GoddardIsotope}

\begin{figure}[t]
\centering
\includegraphics[width=0.47\textwidth]{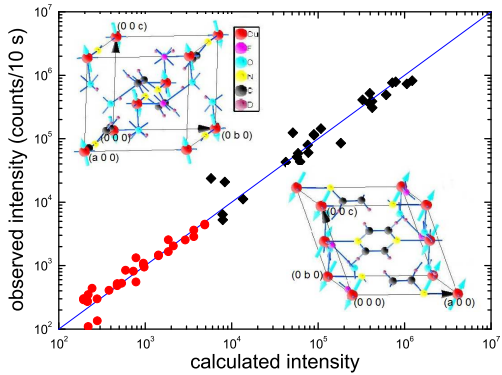}
\caption{\label{fig2} The observed intensity plotted as  a function of calculated intensity for
CuF$_2$(D$_2$O)$_2$(d$_4$-pyz) at T=0.25 K. Nuclear and magnetic reflections are denoted by diamonds and circles, respectively.
The solid line is a guide to the eye. The data were collected on HB-1 in different scattering planes. The insets
depict different views of the unit cell. The arrows on Cu atoms indicates the magnetic moment configuration. The 2d magnetic behavior of CuF$_2$(D$_2$O)$_2$(d$_4$-pyz) originates in the bc plane. }
\vspace*{-3.5mm}
\end{figure}

A total of 26 nuclear and 30 magnetic reflections have been collected in $HK$0 and $H$0$L$
scattering planes at T=0.3 K using HB-1.
Representation analysis for the general case of propagation vector (0.5 0 0) in the structure
symmetry $P21/c$ have been performed using SARAh-Representational Analysis.\cite{SARAh} The
results show that there are two possible irreducible representations as shown in table I. Using the basis vectors in
table I, assuming the magnetic moment direction $\vec{M}$=($M_x$ $M_y$ $M_z$)
varies freely, the magnetic reflections can be fitted. For both cases, when $M_y$ $\neq$ 0, the
spins will configure in noncollinear way. Fitting of moment direction was carried out by minimizing
\begin{equation}\label{eq1}
  \chi^2= \frac{1}{N} \sum [\frac{F^{cal}_M}{F^{obs}_M}-\frac{1}{N} \sum(\frac{F^{cal}_M}{F^{obs}_M})]^2
\end{equation}
\\
where N is the number of reflections. Fits considering both representations were performed and the irreducible representation denoted by $\Gamma_1$
yielded the best agreement with the data.  The resulting collinear spin structure has moments along the [0.7 0 1] real space direction.  For simplicity,
the dipole approximation to the Cu$^{2+}$ form factor has been used. This may lead to a systematic error in both the direction and magnitude of the magnetic
moment due to the anisotropy of the d$_{x^2-y^2}$ orbital.\cite{formfactor} We note that the $\Gamma_1$ irreducible representation with M$_y$=0 results in an antiparallel alignment
of the spins on Cu site 1 (atomic coordinate (0 0 0)) and Cu site 2 (atomic coordinate (0 0.5 0.5)) consistent with the observed propagation vector which
requires K+L=odd.  The resulting magnetic structure for CuF$_2$(H$_2$O)$_2$(pyz), shown in Fig. 2, is G-type (nearest neighbor) antiferromagnetic order.

\begin{table}[htp]
\caption{\label{tab:table} Basis vectors for Cu site 1 (atomic coordinates (0 0 0)) and Cu site 2 (atomic coordinates (0 0.5 0.5)) determined
from the representational analysis\cite{SARAh} for space group No. 14 (P2$_1$/c) and magnetic
propagation vector (0.5 0 0). Here IR represents irreducible representations, BV represents basis vectors.}
\begin{ruledtabular}
\begin{tabular}[b]{cccc}
IR  & BV &  Cu site 1  &  Cu site 2 \\
\hline
$\Gamma_1$ &  $\psi_1$ &  (1 0 0) & (-1 0 0)\\
           &  $\psi_2$ &  (0 1 0) & (0 1 0)\\
           &  $\psi_3$ &  (0 0 1) & (0 0 -1)\\
\hline
$\Gamma_3$ &  $\psi_4$ &  (1 0 0) & (1 0 0)\\
           &  $\psi_5$ &  (0 1 0) & (0 -1 0)\\
           &  $\psi_6$ &  (0 0 1) & (0 0 1)\\
\end{tabular}
\vspace{-2mm}
\end{ruledtabular}
\end{table}

Interestingly the moment direction [0.7 0 1] in real space does not appear to be along an obvious
structural direction. The selection of a specific moment direction suggests some anisotropy in the
spin Hamiltonian.  

The absolute value of the magnetic moment M was determined by comparing the intensity of magnetic and nuclear Bragg reflections. To carefully estimate the moment size, the instrument
resolution was taken into account using Reslib\cite{reslib} and the Debye-Waller factor
has been included. Our calculations yield an ordered moment of 0.60 $\pm$ 0.03 $\mu_B$/Cu.
This value is much smaller than the spin-$\frac{1}{2}$ free ion moment. In low dimensional systems
a moment reduction occurs due to quantum fluctuations and has been observed in a number of
materials.\cite{Tsyrulin,Lumsden} The observed and calculated intensities of magnetic
and nuclear reflections together with the resulting spin arrangement are presented in Fig. 2.

\subsection{spin dynamics}
\begin{figure}[h]
\centering
\includegraphics[width=0.5\textwidth]{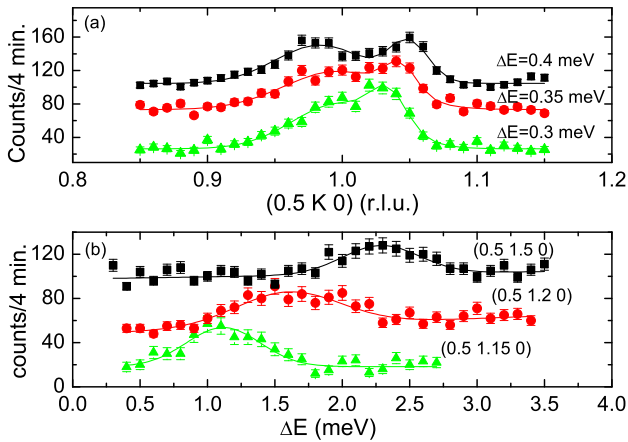}
\caption{\label{fig3a} (a) Constant-$E$ scans along (0.5 $K$ 0) collected in the $HK$0 scattering plane
at T = 1.5 K by the CG-4C spectrometer. (b) Constant-$Q$
scans collected in the $HK$0 scattering plane at T = 1.5 K by the CG-4C spectrometer. The solid lines in both panels are
fits with Gaussians. }
\vspace*{-3.5mm}
\end{figure}


\begin{figure*}[t]
\centering
\includegraphics[width=1\textwidth]{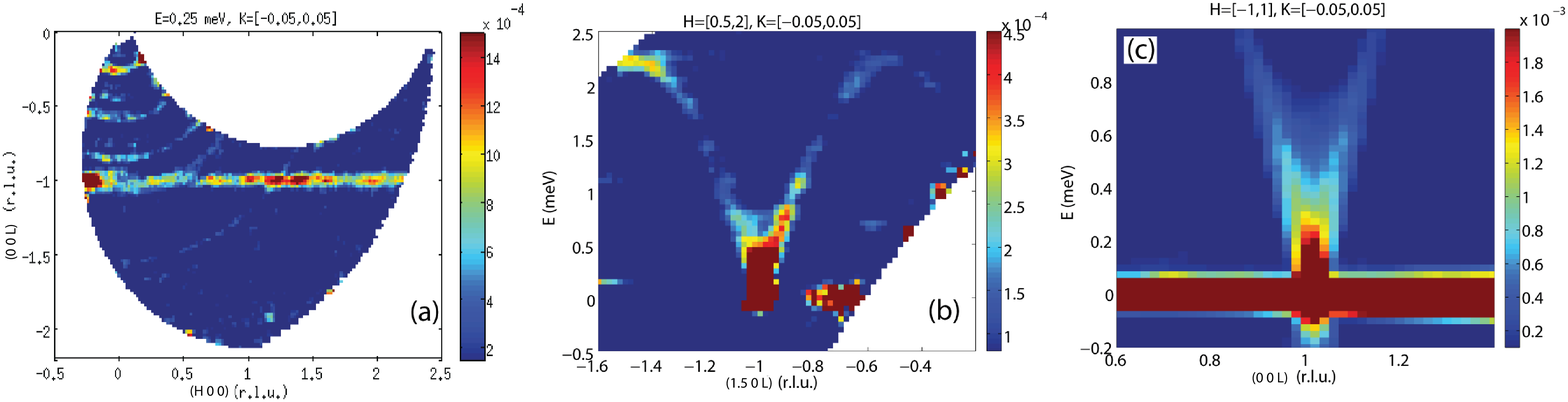}
\caption{\label{fig3} Inelastic neutron scattering data from the time-of-flight spectrometer CNCS. (a) Contour map of intensity in the $H$0$L$ scattering plane with
$\Delta E$=0.25 meV and T = 1.5 K. Background scattering determined from an empty sample holder measurement has been subtracted.
(b) The spin wave dispersion along the (1.5 0 $L$) direction. The data were
collected at T = 1.5 K with $E_i$ = 3 meV. Background scattering determined from an empty sample holder measurement has been subtracted. (c) Low energy spin wave dispersion along the (0 0 $L$) direction in high resolution mode at T = 1.5 K with $E_i$ = 1.5 meV.  Intensities in all panels are given in arbitrary units. The integration range is indicated along with the specified direction at the top of each panel. Note that the dispersion was found to be independent of the value of H an as such the integration range over H was chosen to be large to increase statistics.}
\vspace*{-3.5mm}
\end{figure*}

The spin wave dispersion has been extracted from inelastic neutron scattering measurements.
We first discuss inelastic neutron scattering data collected using the triple-axis spectrometer CG-4C in the $H$$K$0 scattering plane.
Both constant-$Q$ and constant-$E$ scans in the $HK$0 scattering plane have been performed. Constant-$E$ scans
at several different energy transfer are plotted in Fig. 3(a) and constant-$Q$ scans are
shown in Fig. 3(b). These measurements were carried out along the (0.5 $K$ 0)
direction where K=1 corresponds to the magnetic zone center.
The excitation energy
at the zone boundary point $Q$=(0.5 1.5 0) is $E_{zbK}$=2.24$\pm$0.04 meV. Within
instrumental resolution (FWHM=0.17$\pm$0.01 meV), no energy gap at the antiferromagnetic zone
center has been observed. 




\begin{figure}[h]
\centering
\includegraphics[width=0.9\columnwidth]{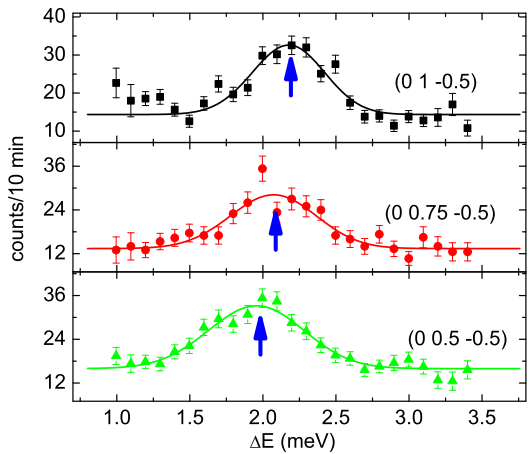}
\caption{\label{fig4} Constant-Q scans along the zone boundary points ranging from
(0 1 -0.5) to (0 0.5 -0.5). The solid line are fits to a Gaussian. the data were
collected on HB-1A. }
\vspace*{-3.5mm}
\end{figure}

In Figs. 4(a), (b) and (c) the spin wave dynamics have been studied with the sample aligned in the
$H$0$L$ scattering plane with the time-of-flight spectrometer CNCS. Fig. 4(a) shows that the spin wave excitation is essentially independent of $H$ at the excitation energy $\Delta E$=0.25 meV, implying only weak interactions along $H$. In Fig. 4(b), the full dispersion
along the $L$-direction is shown. The zone boundary energy in this direction is $E_{zbL}$=2.204$\pm$0.003 meV.
This value is consistent with the previously determined zone boundary energy along the $K$-direction
($E_{zbK}$). To search for a small energy gap, measurements were performed on CNCS
using an incident energy of 1.5 meV resulting in an energy resolution of FWHM=0.034$\pm$0.0007 meV). The measurements were centered on the (0.5 0 1) magnetic zone center.
The data obtained along the $L$-direction is shown in Fig. 4(c). Again, within instrumental resolution (FWHM=0.034 meV),
no energy gap in $L$ dispersion curve was observed. Furthermore, no dispersion was observed along the $H$-direction for L=1. We note that previous DFT calculations predict a
inter-layer exchange constant of 0.011 meV \cite{CuF2Manson} and the resulting interlayer dispersion could not be
observed with our instrumental resolution. Nearly identical exchange interactions along the $K$ and $L$ directions together
with the lack of dispersion along the $H$-direction indicates the magnetism in CuF$_2$(D$_2$O)$_2$(d$_4$-pyz)
is predominately 2d.



The inter-layer $H$-direction magnetic exchange was also probed with
time-of-flight spectrometer CNCS in a high resolution mode with $E_i$ = 1.5 meV. The $H$ vs. $E$
slice, for L=1 (not shown here) is featureless and no obvious spin excitation has been observed within
instrument resolution. Due to the fact that in spin wave theory the zone boundary energy of the spin wave
excitation is of order 2$J$, our data show that the inter-layer exchange interaction $J_{perp}$
is expected to be smaller than $\frac{0.034}{2}$ meV = 0.017 meV. Considering the $E_{zbL}$=2.204 meV
which gives the intra-layer $J$ of 1.102 meV, it is reasonable to estimate that the inter-layer
exchange interaction is less than 1.5\% of the intra-layer exchange interaction. This ratio is in agreement with that determined by both DFT
calculations and by fitting a Heisenberg square lattice model to the magnetic susceptibility which in both cases yields $J_{perp}$/$J_{2d}$ $\sim$ 1\%.\cite{CuF2Manson}

Since no energy gap has been observed in the 2d square lattice plane at the magnetic zone center within
instrumental resolution it is reasonable to believe that any spin exchange anisotropy
must be very small. However, the magnetic moments select the [0.7 0 1] direction suggesting nevertheless that a spin anisotropy is present. As Cu$^{2+}$ moments should have no single ion anisotropy and the
presence of inversion symmetry prohibits Dzyaloshinskii-Moriya interactions, we have included exchange anisotropy
of the following form:
\begin{equation}\label{eq2}
\hat{H}= J \sum [ S^z_i \cdot S^z_j+\Delta (S^x_i \cdot S^x_j+S^y_i \cdot S^y_j)]
\end{equation}
where the summation is over the nearest neighbors in the 2d plane, J is the effective
intra-layer exchange parameter and $\Delta$ is the exchange anisotropy parameter.


Linear spin-wave theory yields the spin wave dispersion
\begin{equation}\label{eq3}
\hbar \omega _Q = 2 J \sqrt{1-\Delta^2 \cos^2(K \pi) \cos^2(L \pi)}
\end{equation}
where $Q$ is a function of K and L.
The dashed lines in the inset of Fig. 6 indicate the diamond pattern on which the Cu-ions are positioned.  Under the approximation that the
diamond is replaced by a square, the abscissa labels in Fig. 6 are related to square lattice notation in the following way: (0.5 1 0) and (0.5 0 1) correspond to ($\pi ~\pi$); (0.5 1.5 0) and (0.5 0 1.5) to ($\pi$/2 $\pi$/2)); and finally the point (0 0.5 -0.5) to ($\pi$ 0).
In Fig. 6 we plot the extracted dispersion relation for CuF$_2$(D$_2$O)$_2$(d$_4$-pyz).
The experimental data in the 2d plane were determined from combinations of triple-axis and time-of-flight
measurements. The solid line in Fig. 6 is the expected dispersion behavior of the classical
Heisenberg model ($\Delta$ = 1) with $J$= 1.102 $\pm$ 0.003 meV.

In the S=$\frac{1}{2}$ Heisenberg square lattice antiferromagnet, it is well established that the spin dynamics
can be well described by the classical linear spin wave theory with the
inclusion of quantum corrections.\cite{Singh,expansion1,expansion2,MontCarlo1,MontCarlo2,Canali}
The net effect of quantum corrections is an overall renormalization factor,
$Z_c \approx$ 1.18, resulting in an effective coupling
constant $J_{eff}$ = $Z_c J_{2d}$ when considering the dispersion from the ($\pi$ $\pi$) zone center
to the ($\pi$/2 $\pi$/2) zone boundary. Thus, our extracted exchange constant, $J$,
is an effective coupling constant and the resulting $J_{2d}$ is 0.934 $\pm$ 0.0025 meV  within the square lattice approximation. This value is
consistent with estimates from the magnetic susceptibility and magnetization
measurements of fully deuterated samples where $J_{2d}$ is about 0.94 meV.\cite{GoddardIsotope}

The dispersion resulting from the anisotropic exchange yields a zone center energy gap of
$E_{zc}$ = $2 J \sqrt{1-\Delta^2}$.  As noted previously, we have not observed such a gap within the
instrumental resolution of 0.034 meV.  This allows us to place a lower bound on $\Delta$.  The coupling constant
extracted from the zone boundary measurements is $J$= 1.102 $\pm$ 0.003 meV.  Placing an upper bound on the gap energy
of 0.034 meV results in a value of $\Delta$ (the ratio $J_{xy} / J_z$) of at least 0.99988. The result would be a very
small anisotropy resulting in a slightly larger $J_z$ which would favor spin orientation along the $z$-axis.
Note that the $z$-axis is the spin wave quantization direction, in this case the real-space [0.7 0 1] direction.  This same
spin anisotropy may be related to the observed low-field spin-flop like transition observed for CuF$_2$(H$_2$O)$_2$(pyz) \cite{CuF2Manson}

The spin wave dispersion has been examined along the antiferromagnetic zone boundary.
In Fig. 5 we present constant-Q scans along (0 $K$ -0.5) from $K$=1 (
square lattice ($\pi$/2 $\pi$/2)) to $K$=0.5 (square lattice ($\pi$ 0)).
In Fig. 6 the observed dispersion between (0 1 -0.5) and (0 0.5 -0.5) has been
plotted. Compared to (0 1 -0.5), the excitation energy at
(0 0.5 -0.5) has been suppressed by 10.3$\pm$1.4\%. This observation is inconsistent with
linear spin wave theory where there is no dispersion along the zone boundary.
However, proper inclusion of quantum fluctuations using series expansion and quantum Monte
Carlo techniques for
the S=1/2 2d AFM square lattice predict a dispersion along the zone boundary with the energy
at ($\pi$ 0) 7-10\% lower than that at ($\pi/$2 $\pi$/2).\cite{Singh,expansion2,MontCarlo2}
Experimental behavior consistent with these calculations have been seen in several model
systems.\cite{Kim,Ronnow,Coldea,Tsyrulin} However, there are some materials where conflicting
experimental observations exist.  For instance, in La$_2$CuO$_4$\cite{Coldea} the energy scale at
($\pi$ 0) is about 7\% larger than
the value at ($\pi$/2 $\pi$/2) in contrast to the expected quantum
behavior which has been attributed to ring exchange among four
spins in the CuO$_2$ plane.
In the coordination polymer
compound Cu(pyz)$_2$(ClO$_4$)$_2$,\cite{Tsyrulin} the excitation energy at ($\pi$ 0) is reduced by 11.5\%
compared to the energy at ($\pi$/2 $\pi$/2). The authors believe this value is larger than the
expectation of series expansion and quantum Monte Carlo calculations and they attribute this stronger
suppression to the next-nearest-neighbor interactions which enhances quantum fluctuations. In
K$_2$V$_3$O$_8$\cite{Lumsden} a striking feature of two modes near the zone boundary
point ($\pi$/2 $\pi$/2) has been observed experimentally and no fully satisfactory explanation has yet been given for this
behavior. For our measurements on CuF$_2$(D$_2$O)$_2$(d$_4$-pyz), in Fig. 6 the dash-dotted line is the
results of calculations including quantum fluctuations with nearest neighbor interactions\cite{expansion2}. It
is clear that the data points follow the expected theory and,
therefore, we interpret the zone boundary dispersion behavior in CuF$_2$(D$_2$O)$_2$(d$_4$-pyz) as being
consistent with the S=1/2 near neighbor 2d AFM square lattice with inclusion of quantum corrections.

\begin{figure}
\centering
\includegraphics[width=0.45\textwidth]{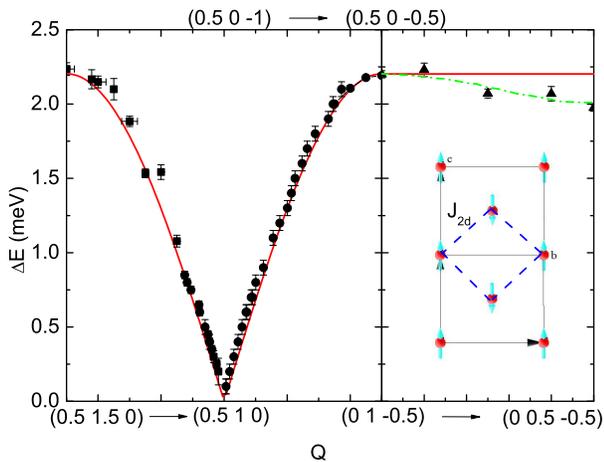}
\caption{\label{fig5} Summary of the spin wave dispersion for CuF$_2$(D$_2$O)$_2$(d$_4$-pyz). The data
were collected using CNCS (solid circles), CG-4C (solid squares) and HB-1A (solid triangles).
The solid line represents the dispersion of Heisenberg linear spin wave theory with a nearest
neighbor interaction. The dash-dotted line represents the results of series expansion to higher order.\cite{expansion2}
The inset depicts the bc-plane. The dashed lines show the square lattice with
a nearest neighbor exchange interaction.
}
\vspace*{-3.5mm}
\end{figure}

\section{Conclusion}

We have performed a series of elastic and inelastic neutron scattering experiments
to study the 2d spin-$\frac{1}{2}$ quasi square lattice antiferromagnet CuF$_2$(D$_2$O)$_2$(d$_4$-pyz).
The 2d magnetic lattice in the bc-plane is the same size as nuclear structure lattice while along the
$a$ direction the magnetic lattice is doubled. The ordered magnetic
moment of the Cu$^{2+}$ ions  is found to be 0.60 $\pm$ 0.03 $\mu_B$ which is significantly
reduced from the expected 1 $\mu_B$, indicating strong quantum fluctuations. The
spins adopt a collinear antiferromagnetic alignment for nearest neighbor sites in the 2d plane and
are oriented along the real-space [0.7 0 1] direction.

The spin dynamics of CuF$_2$(D$_2$O)$_2$(d$_4$-pyz) have been studied using inelastic
neutron scattering data obtained from both triple-axis and time-of-flight instruments. The
2d spin
wave dispersion can be described by a nearly isotropic spin Hamiltonian with small nearest
neighbor interaction $J_{2d}$ = 0.934 $\pm$ 0.0025 meV and very weak inter-layer coupling. 
Along the magnetic
zone boundary, the excitation energy shows a 10.3$\pm$1.4\% dispersion consistent with square lattice calculations
including quantum fluctuations.

\section{Acknowledgments}

Research Work at ORNL was sponsored by the Laboratory Directed Research
and Development Program of ORNL, and was supported by the
Scientific User Facilities Division, Office of Basic Energy Sciences,
DOE. This research was sponsored by the Division of Materials Science
and Engineering of the U.S. Department of Energy (RSF).
This work was supported by UChicago Argonne, LLC, Operator of Argonne National Laboratory (“Argonne”). Argonne, a U.S. Department of Energy Office of Science laboratory, is operated under Contract No. DE-AC02-06CH11357.  Work at EWU was supported by the NSF under grant No. DMR-1005825.

\end{document}